\documentclass[preprint]{elsarticle}
\usepackage{multicol}
\usepackage{color}
\usepackage{xspace}
\usepackage{enumerate}

\begin{document}
\title{Anti-electron Neutrino Event Selection from Backgrounds Based on Machine Learning}

\author[1]{Chang Dong Shin} 
\author[1]{Kyung Kwang Joo\corref{cor1}}
\ead{kyungkwangjoo@gmail.com} 
\author[1]{Dong Ho Moon\corref{cor1}}
\ead{dhmoon@chonnam.ac.kr}

\author[2]{June Ho Choi}
\author[2]{Myoung Youl Pac\corref{cor1}}
\ead{pac@dsu.ac.kr}

\author[3]{Junghwan Goh\corref{cor1}}
\ead{jhgoh@khu.ac.kr}

\cortext[cor1]{Corresponding author}

\address[1]{Institute for Universe \& Elementary Particles, Department of Physics, Chonnam National University, Gwangju 61186, Korea}
\address[2]{Institute for High Energy Physics, Dongshin University, Naju 58245, Korea}
\address[3]{Department of Physics, Kyung Hee University, Seoul 02447, Korea}

\begin{abstract}
For reactor neutrino experiments including the next--generation experiments will be adopting the liquid scintillator technique, criteria and time to select neutrino--induced inverse beta decay events from the background events need to be established. For higher performance efficiency, we investigated the results of applying a machine learning technique embedded in a standard ROOT package to select IBD signals. To obtain a higher statistics, the signals and background events in a gadolinium-loaded liquid scintillation detector were reproduced by Monte Carlo simulation. We report the efficiencies of neutrino--induced $n-H$ and $n-Gd$ events selection using the machine learning technique.
\end{abstract}

\maketitle

\section{Introduction}
Currently, research in particle physics is advancing to address a  large number of physical events to discover new phenomena. Methods for selecting the concerned signals in accordance with the situation and efficient removal of background events have also been newly developed. Machine learning is newly introduced in the field of experimental high energy physics such as LHC, and it is expected to be used for data analysis in other physical fieldss as well\cite{radovic}.

In addition, neutrino, one of the basic particles in the standard model, has been actively researched recently, and the oscillation parameters, in particular, are measured very precisely. However, direct or indirect observational results indicate that there can be new physics beyond the standard model of symmetry breaking in the well-known area including mass sequence of neutrinos, and experiments to determine the presence of inactive neutrinos\cite{ajimura,an,pac}.  In order to unveil these new facts, more precise measurements of the previously measured values ​​are needed. This situation leads us to plan for new experiment to investigate available clues.  Moreover, it is very important to establish a technique for selecting a small amount of signals from a number of background events due to a small scattering cross-sectional area. This is why the neutrino experiments need a large number of neutrino statistics.

In this study, we investigated neutrino signal acquisition while discriminating the background event  through ROOT embedded machine learning tools. Section 2 provides a brief overview of the  analysis setup based on a machine learning toolkit used in this study.  We describe an anti-electron neutrino,$\bar{\nu}_e$ induced inverse beta decay (IBD), background event characteristics in a reactor--based neutrino experiment, Monte Carlo generation schemes and the machine learning tool kit.  Finally, in Section 3, we summarize the results of the neutrino event selection efficiencies and their applications to neutrino experiments which will be launched in near future.    

\section{Analysis Setup}
\subsection{Neutrino Signal Events: Inverse Beta Decay}
In general, a reactor neutrino experiment detects reactor $\bar{\nu}_e$ through the IBD reaction, $\bar{\nu}_e~+~p ~\rightarrow~e^+ ~+~n.$, using liquid scintillator (LS) with 0.1\% gadolinium (Gd) as the target. In the IBD reaction $\bar{\nu}_e$ with energy higher than 1.81 MeV interacts with a free proton in hydrocarbon LS to produce a positron and a neutron. The positron carries most of the kinetic energy of the incoming $\bar{\nu}_e$ while the neutron carries only about 10 keV.
The positron annihilates immediately to releases 1.02 MeV as two  $\gamma$-rays in addition to its kinetic energy. The neutron after thermalization is captured by Gd with a mean delayed time of $\sim 30~ \rm{\mu s}$ and by hydrogen with $\sim 200~\rm{\mu s}$. An IBD candidate event requires a delayed signal from a neutron capture on Gd following the prompt positron annihilation signal. We call it IBD coincidence hereafter.

\subsection{Neutrino Background Events}
There are ``delayed coincidence'' and ``random coincidence'' background events between the prompt and delayed candidates during the detection of  $\bar{\nu}_e$.
The ``random coincidence" background is due to accidental coincidences from the random association of a prompt-like event due to radioactivity and a delayed-like neutron capture. The prompt-like events are mostly ambient $\gamma$-rays from the radioactivity in the photo--mutilplier tube (PMT) glasses, LS and surrounding rock. Most of the ambient radioactivities generate $\gamma$-rays of low energies below $\sim$ MeV. The delayed-like events arise from neutrons produced by cosmic muons in the surrounding rocks or in the detector.

The delayed coincidence" backgrounds are fast neutrons from outside of inner detector, stopping muon followers, $\beta$-n emitters from cosmic-muon induced $\rm{ ^{9}Li/^{8}He}$ unstable isotopes in the target. Fast neutrons are also produced by cosmic muons traversing the surrounding rock and the detector. An energetic neutron entering the ID can interact in the LS to produce a recoil proton before being captured on Gd. The recoil proton generates scintillation lights mimicking a prompt- like event. The $\rm{ ^{9}Li/^{8}He}$-n emitters are mostly produced by energetic cosmic muons because their production cross sections in carbon increase with muon energy.
However, because of  the delayed coincidence backgrounds from high energy muons, we can reject them easily based on abnormally high energy from the muons. It leads us to ignore the delayed coincidence backgrounds in this study.

\subsection{MC Generation}
In neutrino experiment using $\bar{\nu}_e$ beam, there are many criteria for selecting an IBD candidate, however, only three main variables are applied to select the IBD events in our study, energy from prompt signal, $E$, time between a prompt and delayed signal, $\Delta T$ and distance between the two signals, $\Delta R$. The IBD event is likely to have strong correlation between $\Delta T$ and $\Delta R$. but no correlation in the background.\cite{brkim,ahn} Based on these assumptions, we generated four sets of MC, IBD for neutrino--induced $n-Gd$, $n-H$, and backgrounds including delayed and random coincidence corresponding to the two types of IBD event sets. We used a toy MC based on the GLG4SIM package for $n-Gd$ and $n-H$ event samples and extracted the three variables, $E$,  $\Delta T$ and $\Delta R$, event by event. For background, we randomly generated  $E$, $\Delta T$ and $\Delta R$ based on their characteristics in an LS neutrino detector. Figure 1 shows the MC--generated  $E_{prompt}$  for the IBD samples and the corresponding backgrounds. The prompt energies of backgrounds are generated based on the cuts applied to select the IBD candidates used in the reactor neutrino experiments. Considering that an anti--neutrino--induced IBD event distribution is given by the product of $\bar{\nu}_e$  cross section and reactor $\bar{\nu}_e$ flux, we recognize that MC performs its duty well. The $\Delta _T$ distribution is shown in Fig. 2. As shown in the figure 2, $\bar{\nu}_e$--induce--$n_Gd$ and $n-H$ have strong correlation with prompt and delayed signals while the background has no correlation between the fake prompt and delayed signals. Figure 3 depicts the $\Delta_R$ distribution. A clear difference is evident in $\Delta_R$.
\begin{figure}
\centering
\includegraphics[width=110mm]{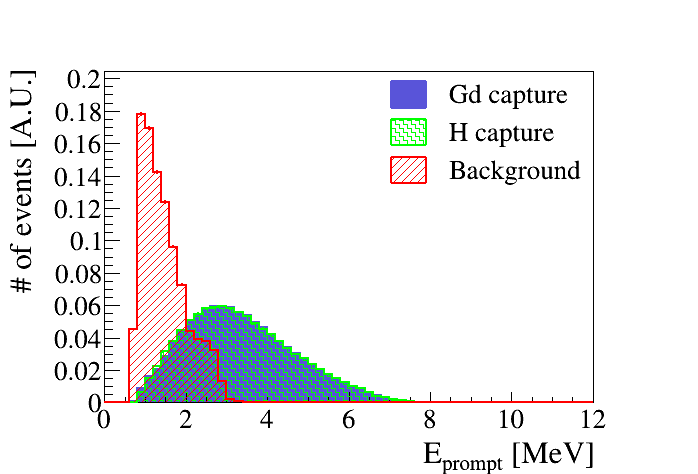}
\caption{Prompt energy distributions from $n-Gd$, $n-H$ and background. There is clear difference between the IBD and background.}
\end{figure}

\begin{figure}
\centering
\includegraphics[width=110mm]{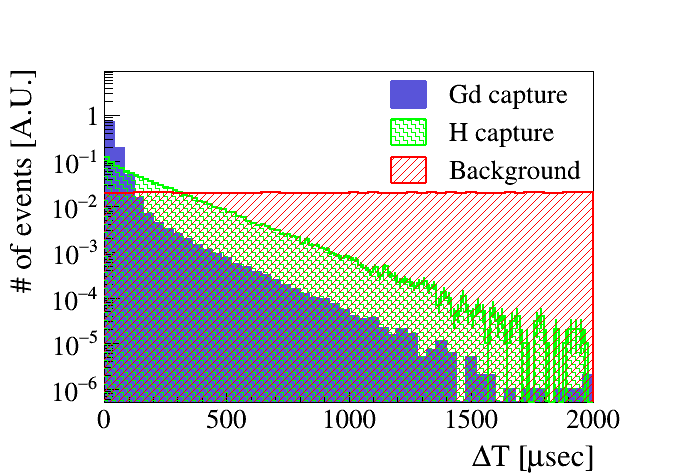}
\caption{Time difference between the prompt and delayed signal, $\Delta T$. Because of neutron cross section on $Gd$ and $H$, two types of IBD show  a different slopes. However, the background has no correlation between the fake prompt and delayed signal. }
\end{figure}
\begin{figure}
\centering
\includegraphics[width=110mm]{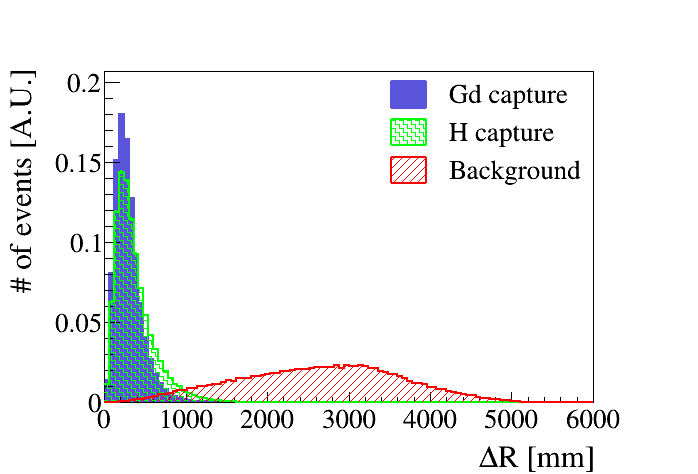}
\caption{Distance between prompt and delayed signal, $\Delta R$. Due to smaller cross section of neutron on $H$, distance between the prompt and delayed signal from $n-H$ is slightly larger than that of $n-Gd$.}
\end{figure}

\subsection{Machine Learning Tool}
The analysis method used in this study is derived from ROOT add-on, namely multilayer perception (MLP)\cite{hoecker}. Neural networks are increasingly being used in various scientific fields for data analysis and classification. The MLP is a simple feed-forward network with the following structure: input layer, first layer of weighting matrix, hidden layer, second layer of weighting matrix, and output layer. It is made up of neurons characterized by a bias and weighted links between them. The input neurons receive the inputs, normalize them and forward them to the first hidden layer. Each neuron in any subsequent layer first computes a linear combination of the outputs of the previous layer. The output of the neuron is then a function of that combination with $f$ being the linear for the output neurons or a sigmoid for the hidden layers. A machine or program is trained with the output $=$ 1 for the signal and 0 for the background, the approximated function of inputs $X$ is the probability of the signal, knowing $X$. Actually, the aim of all the learning methods is to minimize the total error on a set of weighted examples.

\section{Results and Discussion}
The MLP was trained by IBD and the background MC data described in Sec. 3.  To find the most sensitive input variables, 
we created four types of input variable combinations: $E_{prompt}$--$\Delta T$, $E_{prompt}$--$\Delta R$, $\Delta T$--$\Delta R$ and using all variables.

The effectiveness of the MLP learning can be estimated using background rejection as a function of the IBD signal acceptance, namely signal efficiency. Figures 4-7 depict the background rejection efficiencies for these four categories described above. Generally, a neutrino experiment using reactor sourced $\bar{\nu}_e$ reports $\sim$ 75\% of the IBD signal efficiency.  Considering 75\% of IBD selection efficiency, the background rejection efficiencies from the four types of combinations are observed $\geq$98\% at all combinations. Figure 4 shows that the background rejection efficiency using $E_{prompt}$--$\Delta T$ is less effective than the others. In addition, $n-Gd $ is more effective than $n-H$ in Figs. 4-7. This result can be influenced by larger neutron capture cross section on gadolinium than hydrogen.
Larger neutron capture cross sections decrease $\Delta T$ and $\Delta R$. 
From these effects, we can draw the following conclusions: the MLP is likely to consider small $\Delta T$ and short $\Delta R$, which are relevant to the IBD signal.
However, $E_{promp}t$ ls less effective at rejecting the background from the IBD signal. It could due to the fact that we generate the MC based on a predefined background energy, which may dilute the differences between the IBD and background.

However, if we want to reject the background events without distorting the neutrino energy spectrum, we should consider another aspect of the MLP effects related to $E_{prompt}$. Figure 8 and 9 shows the results of $E_{prompt}$ deformation which is defined as the number of events available after background rejection divided by the number of events generated per bin. A flatter line means that more of the neutrino energy spectrum is preserved. In this case, $\Delta T - \Delta R$ combination shows the best results, which could be due to the combination without $E_{prompt}$ This approach can be applied to spectral analysis in neutrino oscillation research with less systematic errors. 

In summary, the effectiveness of the MLP, machine learning, to reject background from the IBD is studied. The MLP can reject background signal as a conventional step-by-step do. Considering a conventional step-by-step cut applying could be more laborious, the MLP approach to neutrino experiment would be a new departure in future.

\begin{figure}
\centering
\includegraphics[width=110mm]{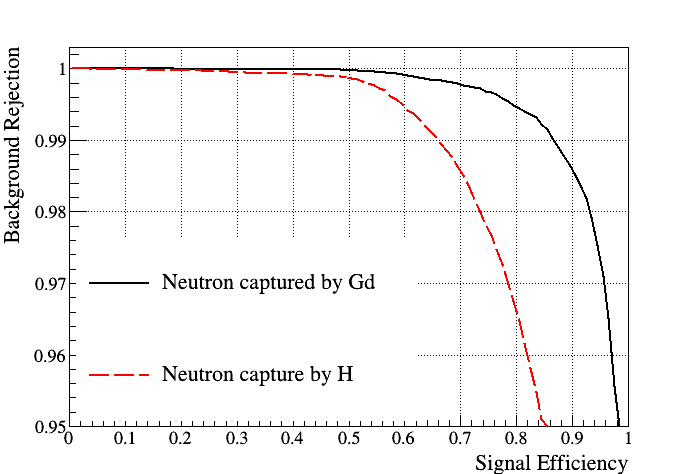}
\caption{Background rejection efficiency by MLP using $E_{prompt}$--$\Delta T$.  Background rejection  from $n-Gd$ is more effective than that from $n-H$.}
\end{figure}

\begin{figure}
\centering
\includegraphics[width=110mm]{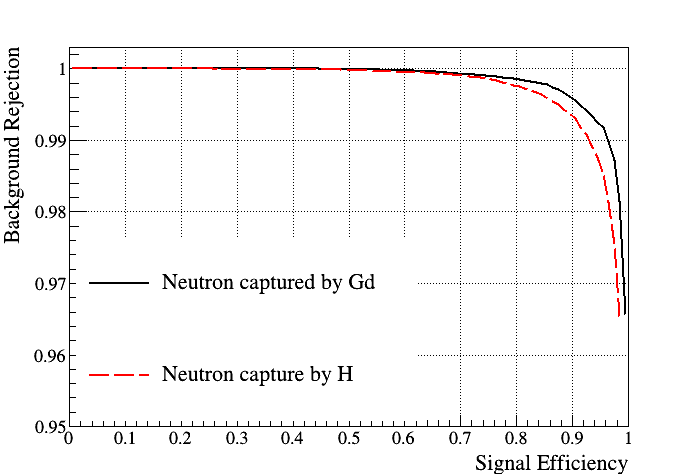}
\caption{Background rejection efficiency by MLP using $E_{prompt}$--$\Delta R$.}
\end{figure}

\begin{figure}
\centering
\includegraphics[width=110mm]{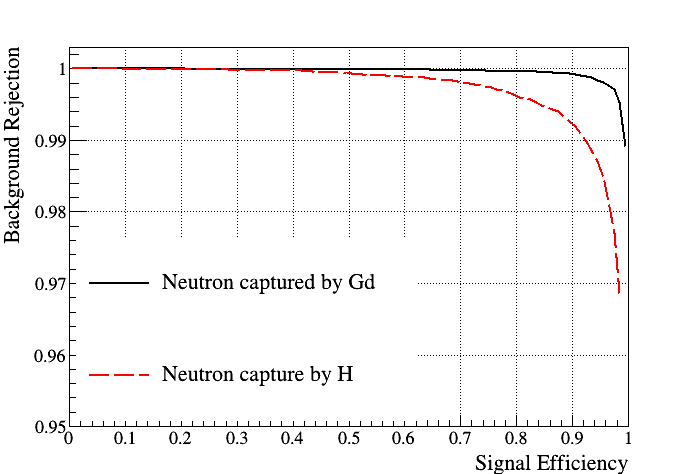}
\caption{Background rejection efficiency by MLP using $\Delta_R$--$\Delta T$.}
\end{figure}

\begin{figure}
\centering
\includegraphics[width=110mm]{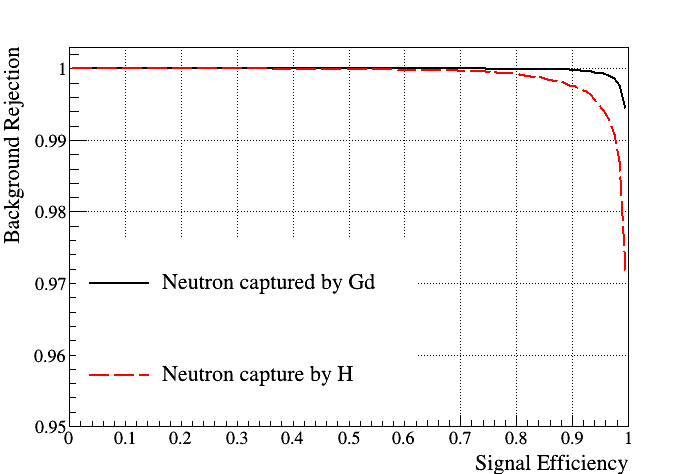}
\caption{Background rejection efficiency by MLP using all variables. The solid line is from $n-Gd$ and the dotted line from $n-H$.}
\end{figure}

\begin{figure}
\centering
\includegraphics[width=110mm]{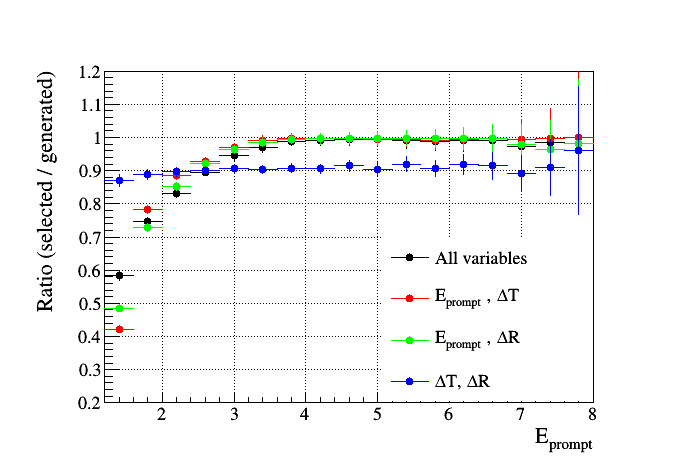}
\caption{Level of the neutrino energy spectrum distortion from $n-Gd$ which is defined as the number of events after background events rejection divided by generated events per bin.}
\end{figure}

\begin{figure}
\centering
\includegraphics[width=110mm]{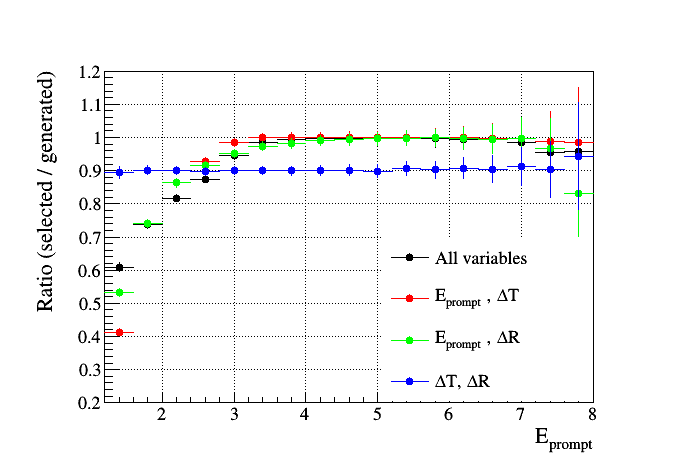}
\caption{Level of the neutrino energy spectrum distortion from $n-H$}
\end{figure}

\section{Acknowledgements}
This work was supported by the National Research Foundation of Korea (NRF-2016R1D1A3B02010606, 2017K1A3A7A09016423, NRF-2018R1A2B6008231, NRF-2019R1I1A1A01059548, and NRF-2019R1A2B5B0107045) and by research fund from  Chonnam National University (2018-3471), Samsung Science \& Technology Foundation (SSTF-BA1402-06) and Kyung Hee University (KHU-20180939).

\end{document}